\documentclass[aps,prd,reprint,twocolumn,supersc riptaddress,nofootinbib]{revtex4-2}
\usepackage{graphicx}
\usepackage{bm}
\usepackage[normalem]{ulem}
\usepackage{color}
\usepackage[colorlinks=true,pdfstartview=FitV,linkcolor=blue,citecolor=blue,urlcolor=blue]{hyperref}
\usepackage[dvipsnames]{xcolor}
\usepackage{amsmath, amsfonts}
\usepackage{booktabs}

\usepackage[caption=false]{subfig} 
\definecolor{bluetto}{HTML}{0088ff}

\definecolor{bluetto}{HTML}{0088ff}

\enlargethispage{\baselineskip}
\everymath{\displaystyle}

\begin{document}

\title{Constraining the axiverse with reionization}

\newcommand{\SMPS}{\affiliation{School of Mathematical and Physical Sciences, University of Sheffield, Hounsfield Road, Sheffield S3 7RH, United Kingdom}}
\newcommand{\KCL}{\affiliation{Theoretical Particle Physics and Cosmology, King’s College London, Strand, London, WC2R 2LS, United Kingdom}}
\newcommand{\TDLI}{\affiliation{Tsung-Dao Lee Institute \& School of Physics and Astronomy, Shanghai Jiao Tong University, Shanghai 201210, China}}
\newcommand{\HARVARD}{\affiliation{Jefferson Physical Laboratory, Harvard University, Cambridge, MA 02138, USA}}

\author{Ziwen Yin}
\email{ziwenyin@sjtu.edu.cn}
\TDLI \KCL
\thanks{ZY and HC contributed equally to this work}

\author{Hanyu Cheng}
\email{chenghanyu@sjtu.edu.cn}
\TDLI \SMPS
\thanks{ZY and HC contributed equally to this work}

\author{Eleonora Di Valentino}
\email{e.divalentino@sheffield.ac.uk}
\SMPS

\author{Naomi Gendler}
\email{ngendler@g.harvard.edu}
\HARVARD

\author{David J. E.\ Marsh}
\email{david.j.marsh@kcl.ac.uk}
\KCL

\author{Luca Visinelli}
\email{lvisinelli@unisa.it}
\affiliation{Dipartimento di Fisica ``E.R.\ Caianiello'', Universit\`a degli Studi di Salerno,\\ Via Giovanni Paolo II, 132 - 84084 Fisciano (SA), Italy}
\affiliation{Istituto Nazionale di Fisica Nucleare - Gruppo Collegato di Salerno - Sezione di Napoli,\\ Via Giovanni Paolo II, 132 - 84084 Fisciano (SA), Italy}

\date{\today}
\begin{abstract}
Axions that couple to electromagnetism are produced in the early Universe by, among other channels, freeze-in via the Primakoff process. For sufficiently large axion masses, the same coupling causes the axions to decay into two photons, which subsequently ionize the intergalactic medium. If this decay occurs in the redshift range $20 \lesssim z \lesssim 1100$, then the contribution to the cosmic microwave background optical depth $\tau_{\rm reio}$ can lead to a conflict with observations, excluding models with sufficiently strongly coupled, heavy axions and high reheating temperatures, $T_{\rm reh}$. 
Using large ensembles of explicit type IIB string theory models with up to $h^{1,1} = 100$ axions, we compute the full cosmic reionization history caused by the decays of multiple axions. We compare this to the posterior on the high-$z$ component of $\tau_{\rm reio}$ derived from parametric-independent constraints on the ionization state of the Universe, obtained in a full \textit{Planck} analysis presented in a companion paper. For $h^{1,1} = 20, 50, 100$, we find that approximately 15\%, 15\%, and 10\% of the models in the ensemble prefer $T_{\rm reh} \lesssim 10^{10}\,\text{GeV}$ at 95\% CL.
We provide a publicly available code at:~\href{https://github.com/ZiwenYin/Reionization-with-multi-axions-decay}{github.com/ZiwenYin/Reionization-with-multi-axions-decay}, which computes the reionization history for arbitrary ensembles of decaying axions. Our analysis opens the door for future large-scale work studying the preference for low-temperature reheating in models with multiple axions.
\end{abstract}

\maketitle


\section{Introduction}
\label{sec:introduction}

The idea that a fundamental pseudoscalar could constitute dark matter (DM) has its origins in the Peccei–Quinn (PQ) solution to the strong-CP problem in QCD~\cite{Peccei:1977hh, Peccei:1977ur}. The spontaneous breaking of this symmetry gives rise to a pseudo-Nambu–Goldstone boson known as the QCD axion~\cite{Weinberg:1977ma, Wilczek:1977pj}. Notably, the QCD axion not only solves the strong-CP problem but also emerges as a compelling DM candidate, provided it is produced through non-thermal mechanisms, such as vacuum misalignment, that allow it to remain cold and non-relativistic well before the epoch of recombination~\cite{Abbott:1982af, Dine:1982ah, Preskill:1982cy} (for reviews, see Refs.~\cite{Marsh:2015xka, DiLuzio:2020wdo, Chadha-Day:2021szb, OHare:2024nmr}). While thermal production via interactions with the Standard Model plasma is also possible, it typically leads to hot or warm relics and is therefore only viable for sufficiently heavy and weakly interacting axion-like particles or axions~\cite{Masso:2002np, Cadamuro:2011fd, Arias:2012az}. Such heavy axions that are abundantly produced by thermal processes are, however, unstable to decay into photons. Thus, thermally produced axions are subject to strong constraints on their parameter space~\cite{Cadamuro:2011fd, Bolliet:2020ofj, Balazs:2022tjl, Langhoff:2022bij, Capozzi:2023xie, Liu:2023nct, Beaufort:2023zuj, Candon:2024eah, Montefalcone:2025nmm}.

String theory compactifications are known to give rise to a large number of axions in low-energy effective field theories (EFTs)~\cite{Witten:1984dg, Svrcek:2006yi, Conlon:2006tq}, spanning a wide range of masses, known as an \emph{axiverse}~\cite{Arvanitaki:2009fg, Acharya:2010zx, Cicoli:2012sz}. The phenomenology of such theories can be rich. In particular, Ref.~\cite{Arvanitaki:2009fg} identified cosmic microwave background (CMB) birefringence, suppressed clustering of DM, black hole superradiance, and axion decays as four key areas of interest. Due to advances in computational algebraic geometry~\cite{Demirtas:2018akl, Demirtas:2022hqf}, it has recently become possible to explore how this phenomenology manifests across large ensembles of explicit compactifications of type IIB string theory. So far, this technology has been deployed to test string theory using suppressed dark matter clustering~\cite{Sheridan:2024vtt}, black hole superradiance~\cite{Mehta:2021pwf}, the strong-CP problem~\cite{Demirtas:2021gsq}, birefringence~\cite{Mehta:2021pwf,Gendler:2023kjt}, and to compute the QCD axion mass as a function of topology~\cite{Gendler:2023kjt, Gendler:2024adn, Benabou:2025kgx}. In this work, we use the method to construct EFTs outlined in Ref.~\cite{Gendler:2023kjt} and subject these models to the rigorous statistical testing we developed in Ref.~\cite{Cheng:2025cmb} from the effects of axion decays on the CMB optical depth.

Using approximate formulae and comparisons to astrophysical limits, it was observed in Ref.~\cite{Gendler:2023kjt} that freeze-in production followed by decay of axions, constrained by the CMB optical depth and other observables~\cite{Langhoff:2022bij, Poulin:2016anj}, seems to place string theory in tension with models of high-temperature reheating. It is the purpose of the present work to place this observation on firm footing with rigorous calculation. Our methodology is the following:
\begin{itemize}
    \item Compute the freeze-in abundance of each individual axion species, using the accurate fitting formulae of Ref.~\cite{Jain:2024dtw} (where it was verified that inter-axion interactions can be neglected).
    \item Compute the full reionization history of the intergalactic medium caused by all\footnote{In practice, we limit to a maximum of ten decaying axions.} axion decays simultaneously, accurately following the baryon temperature, and using energy transfer functions computed separately using \textsc{DarkHistory}~\cite{Liu:2019bbm}.
    \item Compare the resulting contribution to the CMB optical depth to a
    posterior on the high-$z$ ionization
   history reconstructed without assuming any specific parameterization of the reionization history, obtained from the \textit{Planck} low-$\ell$ EE SimAll likelihood, as presented in Paper~I~\cite{Cheng:2025cmb}.
    \item Repeat, varying the reheat temperature $T_{\rm reh}$ to determine the maximum allowed value.
\end{itemize}
We perform this analysis for an ensemble of string theory models described below.

This paper is organized as follows. In Sec.~\ref{sec:axioncosmology}, we review the theoretical framework for axions in string theory, including their production and decay channels. In Sec.~\ref{sec:reionization}, we compute the ionization history of the intergalactic medium resulting from axion decays. We present our model-independent reconstruction of the high-redshift optical depth in Sec.~\ref{sec:highztau}, and compare it to predictions from single-axion models in Sec.~\ref{sec:singleaxion}. In Sec.~\ref{sec:stringconstraints}, we use this framework to constrain ensembles of string theory models. Finally, we summarize our conclusions in Sec.~\ref{sec:conclusions}. The code developed for computing the reionization history with arbitrary populations of decaying axions is publicly available at:~\href{https://github.com/ZiwenYin/Reionization-with-multi-axions-decay}{github.com/ZiwenYin/Reionization-with-multi-axions-decay}, and can be extended to other decaying dark matter scenarios by modifying the relic abundance input.

\section{Axion cosmology}
\label{sec:axioncosmology}

\subsection{Axions in string theory}

Axions arise in the closed-string (gravitational) sector of string theory due to dimensional reduction of $p$-form gauge fields (see Ref.~\cite{Petrossian-Byrne:2025mto} for open-string axions). Consider type IIB string theory, described below the string scale by ten-dimensional supergravity, which contains a 4-form field $C_4$ with field strength $F_5 = {\rm d}C_4$ and action:
\begin{equation}
    S_{\rm 10D} = \int F_5 \wedge \star F_5\, .
\end{equation}

Taking the spacetime manifold to be $M = \mathcal{M}^{3+1} \times X_6$, where $\mathcal{M}^{3+1}$ is Minkowski space with coordinates $x$, and $X_6$ is a Calabi–Yau (CY) orientifold with coordinates $y$, then the low-energy effective field theory below the Kaluza–Klein scale of $X_6$ has $\mathcal{N} = 1$ supersymmetry~\cite{Grimm:2004uq}. The CY topology is partially specified by the Hodge numbers $h^{1,1}$ and $h^{2,1}$, which are split into even and odd components by the orientifold action. For simplicity, consider only the case $h^{1,1} = h^{1,1}_+$, which leads to no orientifold-odd axions (see e.g. Ref.~\cite{Cicoli:2021tzt}).

Taking all four indices of $C_4$ to point along the coordinates in $X_6$, vacuum solutions of the equations of motion for $C_4$ can be found by decomposing in terms of harmonic 4-forms $\omega_i(y)$:
\begin{equation}
    C_4 = \sum_i a_i(x)\,\omega_i(y)\, .
    \label{eqn:C4_expansion}
\end{equation}
Under these model-building assumptions, the index $i$ in Eq.~\eqref{eqn:C4_expansion} runs from $1$ to $h^{1,1}$, and the fields $a_i(x)$ give rise to all $h^{1,1}$ axions in the closed-string sector. Dimensionally reducing the 10D action leads to a 4D action for the axions:
\begin{equation}
    S_{\rm 4D} = \int K_{ij}\,\mathrm{d}a_i \wedge \star\,\mathrm{d}a_j\, ,
\end{equation}
where $K_{ij}$ is the K\"ahler metric, which can be derived from the K\"ahler potential $\mathcal{K} = -2\ln\mathcal{V}$, where $\mathcal{V}$ is the volume of $X_6$ in string units.

The harmonic 4-forms $\omega_i$ are associated to divisors (four-dimensional closed cycles) of $X_6$, the volumes of which are specified by the moduli fields $\varrho_i$, which arise from dimensional reduction of the Einstein–Hilbert term, and form chiral superfields paired with the axions $a_i$. We will assume that the moduli are stabilized~\cite{McAllister:2023vgy, McAllister:2024lnt} with large masses at a point in the ``stretched K\"ahler cone'' of $X_6$: the region where all $\varrho_i$ have volume larger than one in string units and the EFT is under control~\cite{Demirtas:2018akl}. 

The axions are massless to all orders in perturbation theory. They acquire mass due to non-perturbative effects; in particular, we consider Euclidean D3-branes wrapping the divisors of $X_6$, which lead to the scalar potential:
\begin{equation}
    V = -\sum_\alpha \Lambda_\alpha^4 \cos (Q_{\alpha i} a_i + \delta_\alpha)\, ,
\end{equation}
where $\Lambda_\alpha^4 \propto e^{-2\pi Q_{\alpha i} \varrho_i}$ (see Ref.~\cite{Sheridan:2024vtt} for a complete definition), $Q_{\alpha i}$ are integer charges specified by the topology of $X_6$ (in the prime toric basis where the kinetic term is $K_{ij}$), and $\delta_\alpha$ are phases. In addition, we include the instanton potential for QCD; QCD is itself modeled as described below. For the effects of the phases, see the discussions in Refs.~\cite{Demirtas:2021gsq, Mehta:2021pwf, Gendler:2023kjt}.

The EFT for the axions can now be found by expanding the cosine potential, and diagonalizing the K\"ahler metric and mass matrix to give canonically normalized fields defined by a mass $m_i^2$ and decay constant $f_i^2$. The K\"ahler metric terms, which specify $f_i^2$, scale as $M_{\rm Pl}^2/\varrho_i^2$, with $M_{\rm Pl}^2 = 1/(8\pi G_N)$ the four-dimensional Planck mass, while the terms in the mass matrix scale as $e^{-2\pi\varrho_i}$. Thus, variations in the values set for the moduli lead to relatively mild hierarchies in the decay constants, which remain at high scales, while the axion masses become spread out over many decades~\cite{Mehta:2021pwf}. Due to the hierarchies in the scales $\Lambda_\alpha$, it is possible to perform the diagonalization and canonical normalization of the axions perturbatively and efficiently~\cite{Demirtas:2021gsq, Gendler:2023kjt}. It was found in Ref.~\cite{Demirtas:2018akl} that the requirement of having all divisor volumes greater than unity has the effect that, as $h^{1,1}$ increases—thus adding more divisors and more constraints—this forces the volume to become larger and some divisors to be very large. This has the effect on the EFT that axion masses and, more importantly, decay constants fall as more axions are added to the spectrum.

In order to search for evidence of axions beyond their gravitational interactions, we must compute their couplings to the Standard Model. In type IIB string theory, the Standard Model gauge group can be realized by wrapping D7-branes on divisors in $X_6$. We use the toy model described in Ref.~\cite{Gendler:2023kjt} and simply select random prime toric divisors for this purpose (which divisors are suitable in reality is related to the choice of orientifold and the application of Gauss' law to the branes).

We choose one divisor to host the $SU(3)_c$ part of the Standard Model, and either the same divisor or an intersecting divisor to host the $SU(2)_L \times U(1)_Y$ part. We then select a point in the stretched K\"ahler cone, given a homogeneous dilation of the tip, such that the divisor hosting $SU(3)_c$ has volume 40 in string units, giving a gauge coupling $\alpha = 1/40$ at the Kaluza–Klein scale, consistent with the known renormalization group running in the Standard Model assuming high-scale (or no) supersymmetry. We also demand that the $SU(2)_L \times U(1)_Y$ divisor have volume less than 120, such that renormalization group running of the electroweak sector is not drastically altered.

As described in Refs.~\cite{Demirtas:2021gsq, Gendler:2023kjt}, it is now possible to identify the QCD axion~\cite{Peccei:1977hh, Weinberg:1977ma, Wilczek:1977pj} that solves the strong-CP problem (for reviews, see Refs.~\cite{Marsh:2015xka, DiLuzio:2020wdo, Chadha-Day:2021szb, OHare:2024nmr}), and to compute the couplings of all the axions to electromagnetism:
\begin{equation}
    \mathcal{L}_{a\gamma\gamma} = -\frac{1}{4} \sum_i g_{a\gamma\gamma}^i \phi_i F_{\mu\nu} \tilde{F}^{\mu\nu}\,,
\end{equation}
where $g_{a\gamma\gamma}^i$ is the coupling of the canonically normalised $i$-th axion mass eigenstate, $\phi_i$  to photons.

We construct CYs following Batyrev's method~\cite{Batyrev:1993oya} for 3-fold hypersurfaces in ambient toric varieties, which are themselves constructed from triangulations of reflexive polytopes in the Kreuzer–Skarke database~\cite{Kreuzer:2000xy}. This process is automated using \textsc{CYTools}~\cite{Demirtas:2022hqf}, which computes the triple intersection numbers necessary for the relevant computations and returns the K\"ahler metric and divisor volumes given a point in moduli space.\footnote{We do not construct explicit orientifolds, although this is now in principle possible following Moritz~\cite{Moritz:2023jdb}, and implemented at scale in Ref.~\cite{Sheridan:2024vtt} at small $h^{1,1}$. We simply assume that a compatible orientifold with $h^{1,1} = h^{1,1}_+$ exists for the given polytope triangulation.}

To assess the effect of CY topology, we sample $50$ polytopes and $10$ triangulations of each for $h^{1,1} = 20, 50, 100$. For every CY thus constructed, we select every suitable prime toric divisor as a candidate to host QCD, as well as a random intersecting divisor for the electroweak sector. The point in moduli space is specified by K\"ahler parameters $\vec{t}$, which are given by $\lambda \vec{t}_0$, where $\vec{t}_0$ is the tip of the stretched K\"ahler cone and $\lambda$ is chosen such that the QCD divisor takes volume 40 in string units. For the K\"ahler cone, we use $\mathcal{K}_\cup$, which approximates the true CY K\"ahler cone as a union of toric K\"ahler cones (see e.g. Ref.~\cite{Mehta:2021pwf}). This improves on the calculation of Ref.~\cite{Gendler:2023kjt}, which used $\mathcal{K}_V$, the K\"ahler cone of the ambient variety. Since $\mathcal{K}_V \subset \mathcal{K}_\cup$, using the larger $\mathcal{K}_\cup$ allows one to include more small divisors. $\mathcal{K}_\cup$ is a better approximation to the true K\"ahler cone of the CY, and the additional small divisors available compared to $\mathcal{K}_V$ allow a more representative sampling of the moduli space. Computational constraints limit using $\mathcal{K}_\cup$ to smaller values of $h^{1,1} \lesssim 100$, and so we could not explore the largest value $h^{1,1} = 491$ as in Ref.~\cite{Gendler:2023kjt}.

\subsection{Axion production and decay}

\begin{figure*}[htbp]
    \centering
    \includegraphics[width=0.49\linewidth]{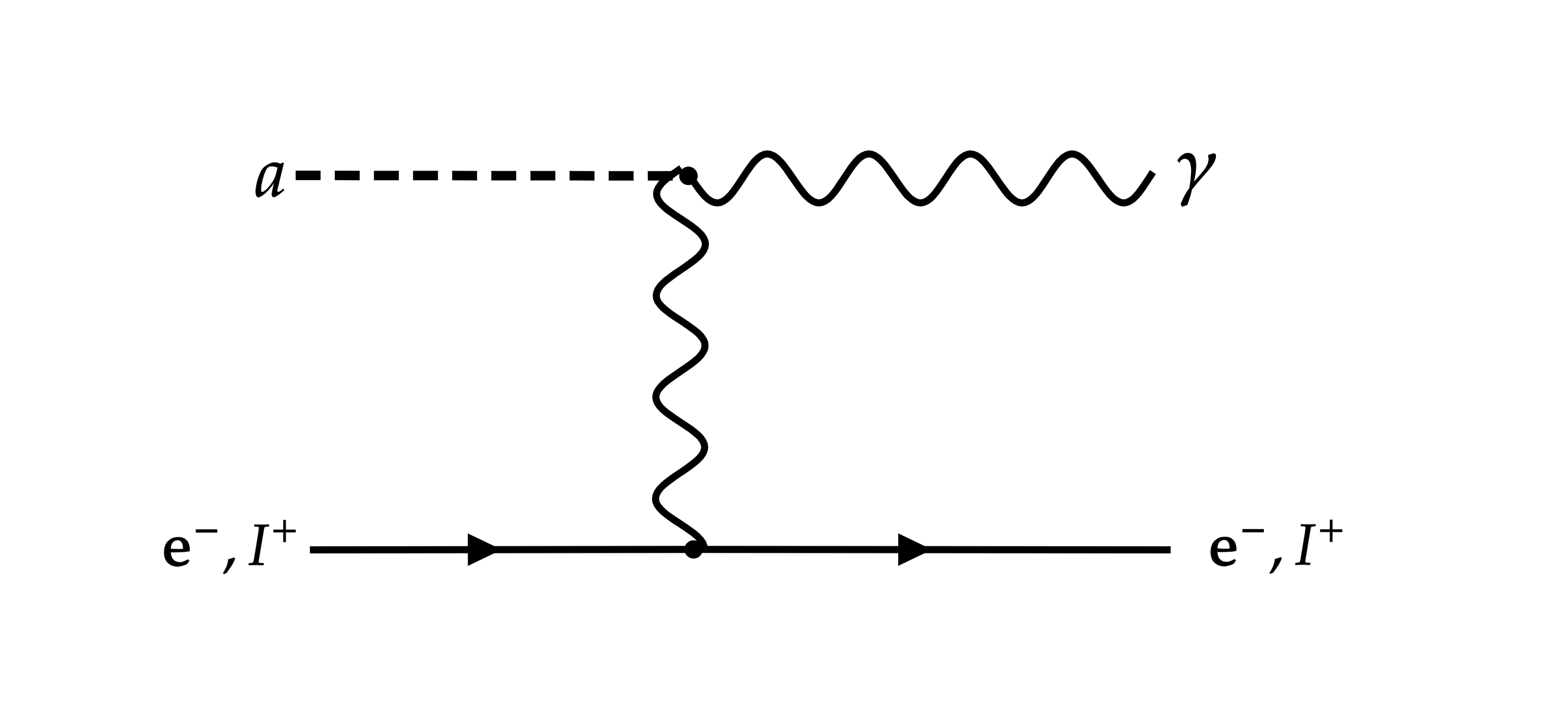}
    \includegraphics[width=0.49\linewidth]{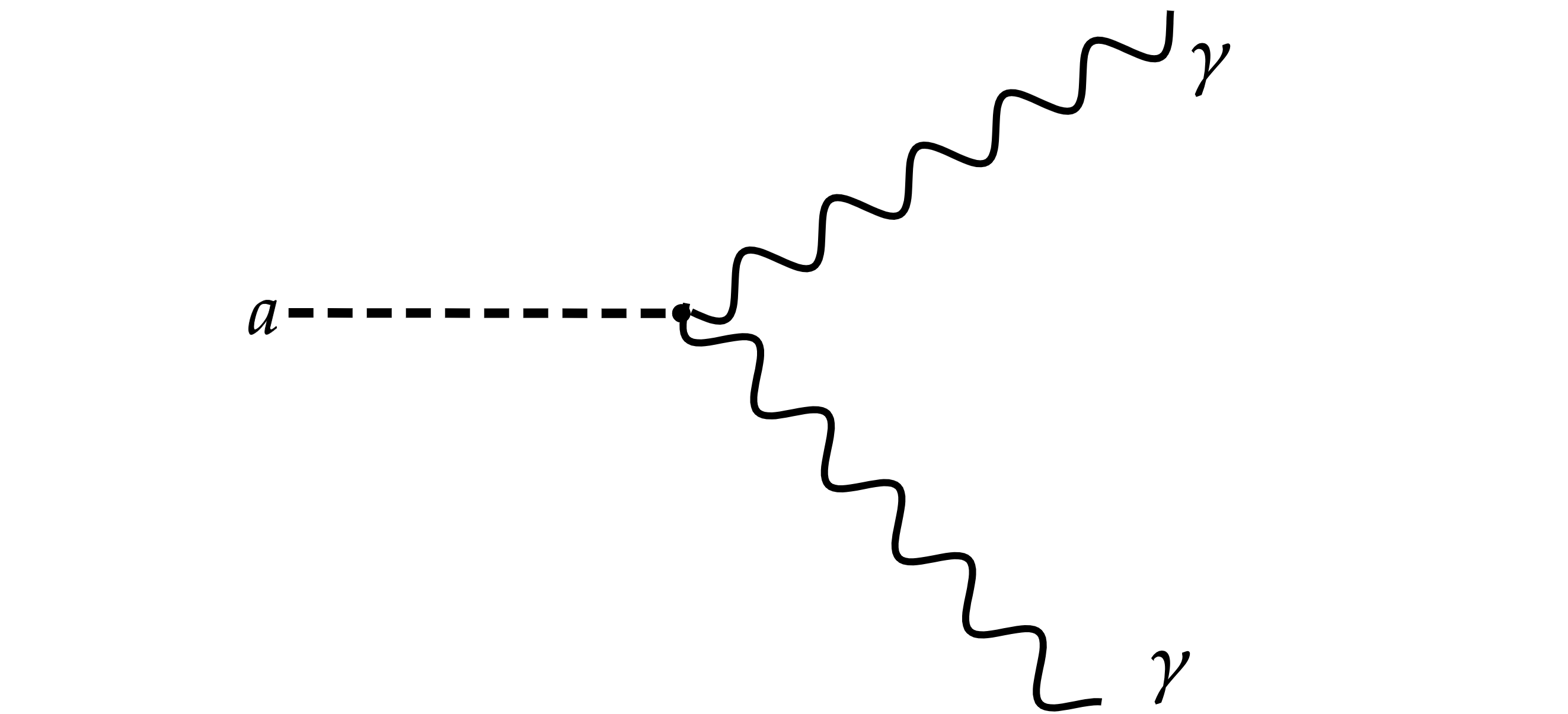}
    \caption{Left: Axion production by the Primakoff process. Right: Axion to two photon decay.}
    \label{fig:primakoff_decay}
\end{figure*}

The axion–photon coupling leads to axion production in the early Universe primarily through the Primakoff process, which dominates when axions couple to photons (we conservatively set the contribution from misalignment to zero). In this process, thermal photons scatter off charged particles in the Standard Model plasma, $q + \gamma \rightarrow q + a$, converting into axions in the presence of electromagnetic fields sourced by these charged particles. The same interaction governs axion decay to two photons, with decay rate:
\begin{equation}
    \label{eq:Gamma}
    \Gamma_{a\rightarrow\gamma\gamma} = \frac{m_a^3 g_{a\gamma\gamma}^2}{64\pi} \, ,
\end{equation}

where in this equation and in the following we refer to individual axions and for simplicity drop the subscript $i$ labeling them. If the axion mass is sufficiently large, $m_a \gg 1\,\text{eV}$, then the decay products leave astrophysical signatures and strongly constrain the axion parameter space. These processes are depicted in Fig.~\ref{fig:primakoff_decay} for the Primakoff production (left) and particle decay (right). Here, we compute the decay temperature $T_{\rm decay}$ and the corresponding redshift $z_{\rm decay}$ using the methods outlined in Ref.~\cite{Jain:2024dtw}.

Early studies of this phenomenology~\cite{Cadamuro:2010cz, Cadamuro:2011fd} made the assumption that the axions came into thermal equilibrium in the early Universe, followed by freeze-out. In this case, constraints on $g_{a\gamma\gamma}$ are incredibly powerful, only allowing $g_{a\gamma\gamma} \ll M_{\rm Pl}^{-1}$. However, the assumption of freeze-out is only valid if the reheating temperature (maximum thermalization temperature), $T_{\rm reh}$, is larger than the freeze-out temperature of the dominant Primakoff interaction~\cite{Cadamuro:2010cz, Cadamuro:2011fd}:
\begin{equation}
    \label{eq:primakoff_rate}
    T_{\rm fo} \approx 10^{16} \frac{\sqrt{g_\star}}{g_{*Q}} \left(\frac{10^{-16}\,\text{GeV}^{-1}}{g_{a\gamma\gamma}}\right)^2\, \text{GeV}\, ,
\end{equation}
where $g_\star$ and $g_{*Q}$ are the relativistic and charged number of degrees of freedom in the plasma, respectively.

For lower reheating temperatures, $T_{\rm reh} < T_{\rm fo}$, axions are instead produced through freeze-in~\cite{Hall:2009bx}, rather than reaching thermal equilibrium. In this regime, the resulting axion abundance depends sensitively on the reheating temperature~\cite{Balazs:2022tjl, Langhoff:2022bij, Jain:2024dtw}, with a minimum set by the ``irreducible'' contribution at big bang nucleosynthesis (BBN), $T_{\rm reh} \gtrsim T_{\rm BBN} \approx 5\,\text{MeV}$ (see e.g.\ Refs.~\cite{deSalas:2015glj, Hasegawa:2019jsa}), and a maximum at $T_{\rm reh} \lesssim 10^{16}\,\text{GeV}$, corresponding to the maximum temperature allowed in inflationary models, with the energy scale constrained by searches for primordial gravitational waves in the CMB~\cite{Planck:2018vyg}.

The freeze-in energy density of axions produced from the Primakoff process can be parameterized as~\cite{Jaeckel:2014qea}
\begin{align}
\label{eq.primfreezein_full}
    &\rho_\mathrm{Prim}(T \ll T_{\mathrm{reh}})\\
    &= A_\mathrm{Prim} \frac{2.4 \times 10^{-8}}{\sqrt{g_{*}(T_\mathrm{reh})}}\!\left(\frac{T_{\mathrm{reh}}}{10\,\mathrm{MeV}}\right)\!\left(\frac{g_{a\gamma\gamma}}{10^{-11}\,\mathrm{GeV^{-1}}}\right)^2\!\left(\frac{m_a}{1\,\mathrm{MeV}}\right)\notag \\
    &\times \left(\frac{a(T_\mathrm{reh})}{a(T)}\right)^3 \left(\frac{n_\mathrm{eq}(T_{\mathrm{reh}})}{1\,\mathrm{MeV}^3}\right)\left(\frac{g_{*Q}(T_{\rm{reh}})}{g_{*Q}(5\,\rm{MeV})}\right)\, \mathrm{MeV}^4\, ,\notag
\end{align}
where $n_{\mathrm{eq}}$ is the number density of axions in thermal equilibrium at $T = T_{\mathrm{reh}}$. In the limit $m_a \ll T_{\mathrm{reh}}$, the expression above simplifies to
\begin{align}
    \label{eq:primakoff}
    & \rho_\mathrm{Prim}(T \ll T_{\mathrm{reh}}) \\
    &= A_\mathrm{Prim} \frac{2.4 \times 10^{-5}}{\sqrt{g_{*}(T_\mathrm{reh})}} \frac{\zeta(3)}{\pi^2} \left(\frac{T_{\mathrm{reh}}}{10\,\mathrm{MeV}}\right)^4 \left(\frac{g_{a\gamma\gamma}}{10^{-11}\,\mathrm{GeV^{-1}}}\right)^2 \notag\\
    &\times \left(\frac{a(T_\mathrm{reh})}{a(T)}\right)^3 \left(\frac{m_a}{1\,\mathrm{MeV}}\right)\left(\frac{g_{*Q}(T_{\rm{reh}})}{g_{*Q}(5\,\rm{MeV})}\right)\, \mathrm{MeV}^4\, ,\notag
\end{align}
where the numerical coefficient $A_\mathrm{Prim}$, which depends on $m_a$ and $T_{\mathrm{reh}}$, is calculated in Ref.~\cite{Jain:2024dtw}. The resulting freeze-in fraction of decaying axion dark matter for $T \gg T_{\rm decay}$ is then
\begin{align}
    \label{eq:primakoff_approx}
    &\mathcal{F}_\mathrm{Prim} 
    \equiv \frac{\rho_\mathrm{Prim}}{\rho_{\mathrm{DM}}} \\
    &= \frac{3.4 \times 10^{-3} A_\mathrm{Prim}}{\sqrt{g_{*}(T_\mathrm{reh})}} \left(\frac{T_{\mathrm{reh}}}{10\,\mathrm{MeV}}\right) \left(\frac{g_{a\gamma\gamma}}{10^{-11}\,\mathrm{GeV^{-1}}}\right)^2 \notag\\
    &\times \left(\frac{m_a}{1\,\mathrm{MeV}}\right)\left(\frac{n_{\rm{eq}}(T_{\rm{reh}})}{T_{\rm{reh}}^3}\right)\left(\frac{g_{*s}(T_{\gamma 0})}{g_{*s}(T_{\rm{reh}})}\right)\left(\frac{g_{*Q}(T_{\rm{reh}})}{g_{*Q}(5\,\rm{MeV})}\right)\, .\notag
\end{align}
where $g_{*s}$ is the effective number of relativistic degrees of freedom for entropy.

A complementary freeze-in channel arises from the inverse decay process $\gamma + \gamma \rightarrow a$, when the kinematical condition $m_a>2m_\gamma$ is satisfied with the resulting relic fraction given by
\begin{align}
\label{eq:inversedecay_approx}
    \mathcal{F}_\mathrm{Id} &= A_\mathrm{Id} \frac{1.42 \times 10^{-4}}{\sqrt{g_*(T')}} \left(\frac{g_{a\gamma\gamma}}{10^{-11}\,\mathrm{GeV}^{-1}}\right)^2 \left(\frac{m_a}{0.1\,\mathrm{MeV}}\right)^2 \notag\\
    &\times \!\left(\!\frac{T'}{T_\mathrm{Id}} \frac{\left(4\left(\frac{T_\mathrm{Id}}{T'}\right)^2 \!-\! 1\right)^{3/2}}{3^{3/2}} \frac{\coth\!{\left(0.1 \frac{T_\mathrm{Id}}{T'}\right)}}{\coth{(0.1)}} \!\right)_{T' = \min[T_\mathrm{Id}, T_\mathrm{reh}]}\!\notag\\ 
&\times\left(\frac{g_{*s}(T_{\gamma 0})}{g_{*s}(T')}\frac{g_{*Q}(T')}{g_{*Q}(5\,\rm{MeV})}\frac{n_{\rm{eq}}(T')}{T'^3}\right)_{T' = \min[T_\mathrm{Id}, T_\mathrm{reh}]} ,
\end{align}
where $T_\mathrm{Id} \simeq 2.5 m_a$ is the inverse decay threshold, $m_\gamma$ is the plasma mass and $A_\mathrm{Id}$ is a phenomenological factor that solely depends on the axion mass~\cite{Jain:2024dtw}. In the region of interest, the contribution from $e^+ e^-$ annihilation is negligible. Finally, we define the total abundance of axions, $\mathcal{F}_a$, “as if the axion were stable”~\cite{Langhoff:2022bij}, as the sum of the freeze-in contributions from the Primakoff and inverse decay processes:
\begin{equation}
    \label{eq:F}
    \mathcal{F}_a \simeq \mathcal{F}_\mathrm{Prim} + \mathcal{F}_\mathrm{Id}\, .
\end{equation}
The above fitting formulae allow us to compute the fraction of would-be dark matter composed of axions before their decay, as a function of the axion mass $m_a$, the axion–photon coupling $g_{a\gamma\gamma}$, and the reheat temperature $T_{\rm reh}$. This fraction serves as the input for the next stage of our analysis.

Note that it is implicitly assumed that such reheating is instantaneous, i.e. inflation ends at $T_{\rm reh}$ and there are no axions produced in the prior epoch. Similarly to neglecting the possible misalignment population of axions (which can be fine-tuned away), the assumption of instantaneous reheating leads to conservative constraints from the irreducible freeze-in population. Alternatively, one could consider explicit modified cosmologies~\cite{Visinelli:2009kt,Sheridan:2024vtt} or take the abundance as an independent free parameter with $\mathcal{F}_a$ above giving the minimum~\cite{Balazs:2022tjl}. 

\section{Ionization of the intergalactic medium by axion decay}
\label{sec:reionization}

We first consider the ionization rate of our Universe between recombination and reionization, corresponding to the redshift range $10\lesssim z\lesssim 1100$. The recombination epoch  is mainly governed by the evolution of the ionization fraction $X_e$, defined as the ratio between the number densities of free electrons $n_e$ and hydrogen nuclei $n_H$,
\begin{equation}
    \label{eq.ionization rate}
    X_e=\frac{n_e}{n_H}\,,
\end{equation}
and the baryon temperature $T_b$. More precisely, the differential equations describing the evolution of these quantities read~\cite{Escudero:2023vgv,Slatyer:2016qyl}
\begin{align}
  &\frac{{\rm d}X_e}{{\rm d}z}\\
  =&\frac{C_H}{(1+z)H(z)}\!\left(\!-\beta_H(T_\gamma)(1 \!-\! X_e)e^{\frac{-E_{H,2s1s}}{T_\gamma}} \!\! +\! X_e^2n_H\alpha_H(T_b)\!\right)\notag\\
  &+\frac{1}{n_HE_i}\frac{{\rm d}E}{{\rm d}V{\rm d}z}\bigg|_{\mathrm{dep,ion}}\!\!\!\!+(1-C_H)\frac{1}{n_HE_{H,1s2p}}\frac{{\rm d}E}{{\rm d}V{\rm d}z}\bigg|_{\mathrm{dep,exc}}\!\!,\notag \\
    &(1+z)\frac{{\rm d}T_b}{{\rm d}z}=2T_b+\frac{8}{3}\frac{\rho_\gamma\sigma_T}{m_e\,H(z)}\frac{X_e}{1+f_{\mathrm{He}}+X_e}(T_b-T_\gamma)\notag\\
    &+\frac{2}{3}\frac{1+z}{n_H(1+f_{\mathrm{He}}+X_e)}\frac{{\rm d}E}{{\rm d}V{\rm d}z}\bigg{|}_\mathrm{dep,heat}\,, 
\end{align}
where $T_\gamma$ is the photon temperature, $H(z)$ is the Hubble parameter at redshift $z$, $E_{H,1s2s}=E_{H,1s2p}=10.2\,\mathrm{eV}$ is the energy difference between the 2s and 1s energy levels of the hydrogen atom, and $E_i=13.6\,\mathrm{eV}$ is the average ionization energy of hydrogen. The factor $C_H$ gives the probability for an electron in the first excited state to transition to the ground state before being ionized~\cite{Slatyer:2016qyl}. Moreover, the evolution of $T_b$ is influenced by the Thomson scattering cross section $\sigma_T$ and the photon energy density $\rho_\gamma$, while the coefficients $\beta_H$ and $\alpha_H$ appearing in the evolution of $X_e$ correspond to photoionization and recombination rates, respectively.

For each energy deposition channel, labeled by ``$\mathrm{c}$'', with ``ion'' for ionization, ``exc'' for excitation, and ``heat'' for IGM heating, the energy deposition rate takes the form
\begin{equation}
    \frac{{\rm d}E}{{\rm d}V\,{\rm d}z}\bigg|_{\mathrm{dep,c}} \equiv f_{\mathrm{c}}(z)\,\frac{{\rm d}E_{\mathrm{inj}}}{{\rm d}V\,{\rm d}z}\,,
\end{equation}
where $f_{\mathrm{c}}(z)$ is the redshift-dependent efficiency for the channel ``$\mathrm{c}$''. In contrast to simplified on-the-spot treatments, where the deposited power is approximated by a simple exponential decay law, we employ the full redshift-dependent energy deposition efficiency $f_{\mathrm c}(z)$ computed using \textsc{DarkHistory}~\cite{Liu:2019bbm}, which consistently account for delayed energy deposition from EM cascades.

The energy injection rate due to axion decays is given by~\cite{Lucca:2019rxf}
\begin{equation}
    \frac{{\rm d}E_{\mathrm{inj}}}{{\rm d}V\,{\rm d}z} = \frac{\Gamma_{a\rightarrow\gamma\gamma} \,e^{-\Gamma_{a\rightarrow\gamma\gamma} t(z)}}{(1+z)\,H(z)}\,\mathcal{F}_a\,\rho_c\,\Omega_{\rm DM}\,(1+z)^3\,,
\end{equation}
where $\Gamma_{a\rightarrow\gamma\gamma}$ is the axion decay rate defined in Eq.~\eqref{eq:Gamma}, $\mathcal{F}_a$ is the DM fraction in axions prior to decay in Eq.~\eqref{eq:F}, $\Omega_{\rm DM}$ is the present DM density fraction, and $\rho_c$ is the critical density today. We also define the total number density of hydrogen, including both neutral ``$H$'' and ionized ``$HI$'', as a function of photon temperature $T_\gamma$:
\begin{equation}
    n_H=n_e+n_{HI}=3.1\times 10^{-8}\eta_{10}(T_\gamma/T_{\gamma0})^3\,\mathrm{cm}^{-3}
\end{equation}
where $\eta_{10}=10^{10}\,n_b/n_\gamma\simeq 6.1$ is the baryon-to-photon ratio~\cite{Fields:2019pfx}, $n_b$ is the baryon number density, and $T_{\gamma0}\approx 2.7$\,K is the present-day photon temperature. This leads to the photon temperature at redshift $z$ as:
\begin{equation}
    T_\gamma=T_{\gamma0}(1+z)\,.
\end{equation}

The reionization era, corresponding to redshifts $z\lesssim 10$, is modeled using the tanh parametrization, see e.g.\ Ref.~\cite{Heinrich:2021ufa}. This describes the reionization of hydrogen and singly ionized helium as:
\begin{equation}
  X_{\mathrm{rei}}=\frac{1+f_{\mathrm{He}}}{2}\left[1+\tanh\left(\frac{(1+z_{\rm re})^{3/2}-(1+z)^{3/2}}{(3/2)(1+z)^{1/2}\Delta z}\right)\right]\,,
\end{equation}
where $z_{\rm re}=6.1$, $\Delta z=0.5$ and the helium-to-hydrogen number density ratio is given by
\begin{equation}
    f_{\mathrm{He}}=\frac{n_{\mathrm{He}}}{n_\mathrm{H}}=\frac{Y_p}{4(1-Y_p)}\,,
\end{equation}
with $Y_p$ denoting the primordial helium mass fraction. In this model, the full ionization history is expressed as the sum of the contributions from recombination and reionization:
\begin{equation}
    X_e^{\mathrm{tot}}(z)=X_e(z)+X_{\mathrm{rei}}(z)\,.
\end{equation}

\begin{figure}[htbp]
    \centering
    \includegraphics[width=1\linewidth]{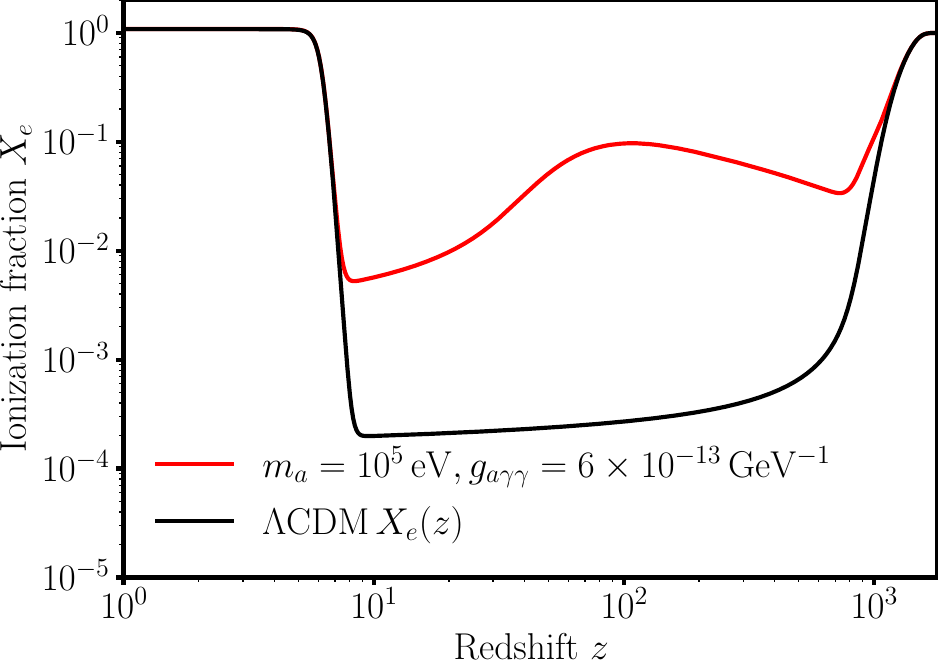}
    \caption{\textbf{Single axion case.} Ionization fraction in the $\Lambda$CDM scenario (black line), compared to a model with one additional axion component and $T_{\rm reh}=5$\,MeV, 
    using the full redshift-dependent energy deposition efficiency computed with \textsc{DarkHistory}.
    The axion, constituting a fraction $\approx 10^{-7}$ of the DM, decays at redshift $z_{\rm decay}<1100$, leading to efficient ionization of the intergalactic medium and a CMB optical depth $\tau_{\mathrm{highz}} \approx 2.24$. }
    \label{fig:Xez}
\end{figure}

To illustrate the impact of axion decays on the ionization history, Fig.~\ref{fig:Xez} shows the evolution of the ionization fraction in a model with a single axion-like particle of mass $m_a = 10^5\,\mathrm{eV}$ and photon coupling $g_{a\gamma\gamma} = 6 \times 10^{-13}\,\mathrm{GeV}^{-1}$, assuming a low reheating temperature $T_{\rm reh} = 5\,\mathrm{MeV}$. The axion population, comprising a small fraction $\sim 10^{-7}$ of the DM, decays after recombination and induces substantial ionization of the intergalactic medium at high redshift, compared to the standard cosmological history (black line). This early energy injection results in an optical depth $\tau_{\mathrm{highz}} \approx 2.24$, which is in strong tension with observational limits.

\begin{figure*}[htbp]
    \centering
    \includegraphics[width=1\linewidth]{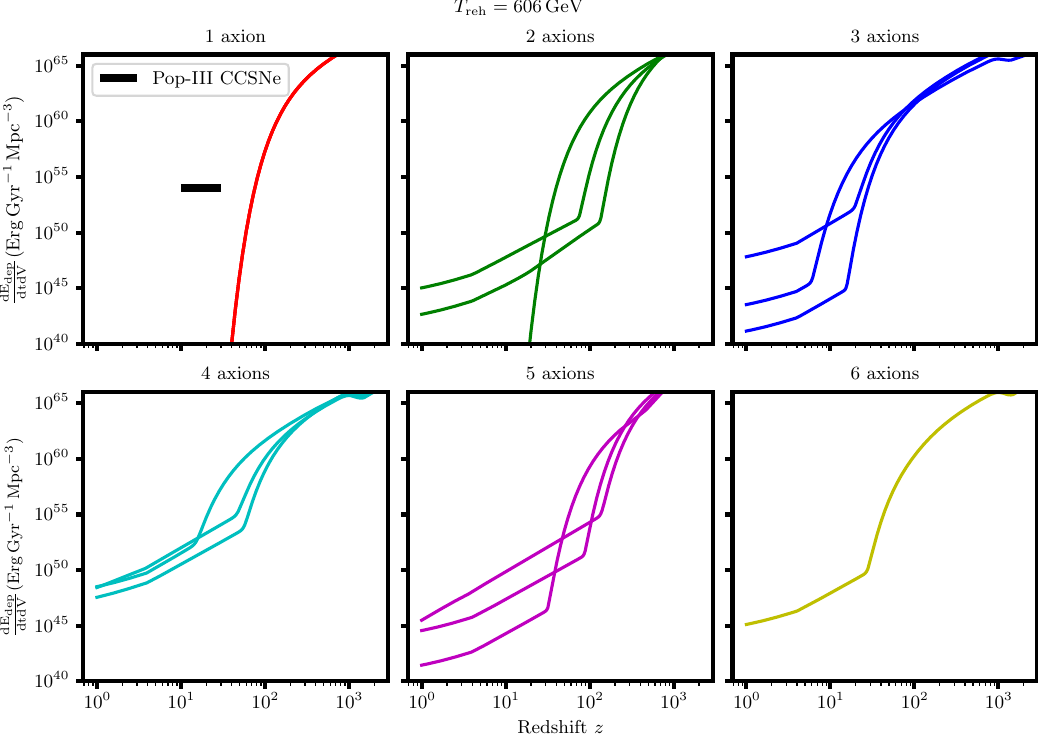}
    \caption{Energy deposition rates from representative models with different numbers of axions in the selected mass window relevant for decay and ionization. In each panel, up to three curves are shown, corresponding to representative models drawn from the $h^{1,1} = 50$ ensemble of Ref.~\cite{Gendler:2023kjt}. Panels with fewer curves reflect the limited number of models in the ensemble realizing those configurations within the chosen mass range. For reference, the black line shows the energy injection rate and redshift range of Population III core collapse supernovae (Pop-III CCSNe)~\cite{Hartwig:2022lon}.}
    \label{fig:energy_injection_h1150}
\end{figure*}

In the string axiverse, multiple axions with hierarchically different masses and decay constants can lead to distinct episodes of energy injection in the post-recombination Universe. To study scenarios with a large number of axions, we focus on a mass window ranging from $10^5\,\mathrm{eV}$ to $10^{10}\,\mathrm{eV}$, and select up to 10 axions per model within this window to compute the cumulative energy injection. This mass range is chosen because, for the typical axion--photon couplings in our string-motivated models, it corresponds to axions that decay in the redshift interval relevant for CMB constraints on reionization, $20 \lesssim z \lesssim 1100$.Figure~\ref{fig:energy_injection_h1150} shows representative energy injection histories from the models presented in Ref.~\cite{Gendler:2023kjt}, sampled from an ensemble of CY compactifications with $h^{1,1} = 50$. The selected cases include models with between one and six axions in the specified mass window. In this ensemble, the maximum number of axions found in the mass window for any given model is 8. In each panel, we display up to three representative models to maintain clarity. Panels with fewer curves reflect the limited number of models in our ensemble realizing those configurations within the chosen mass range.

\begin{figure}[htbp]
    \centering
    \includegraphics[width=1\linewidth]{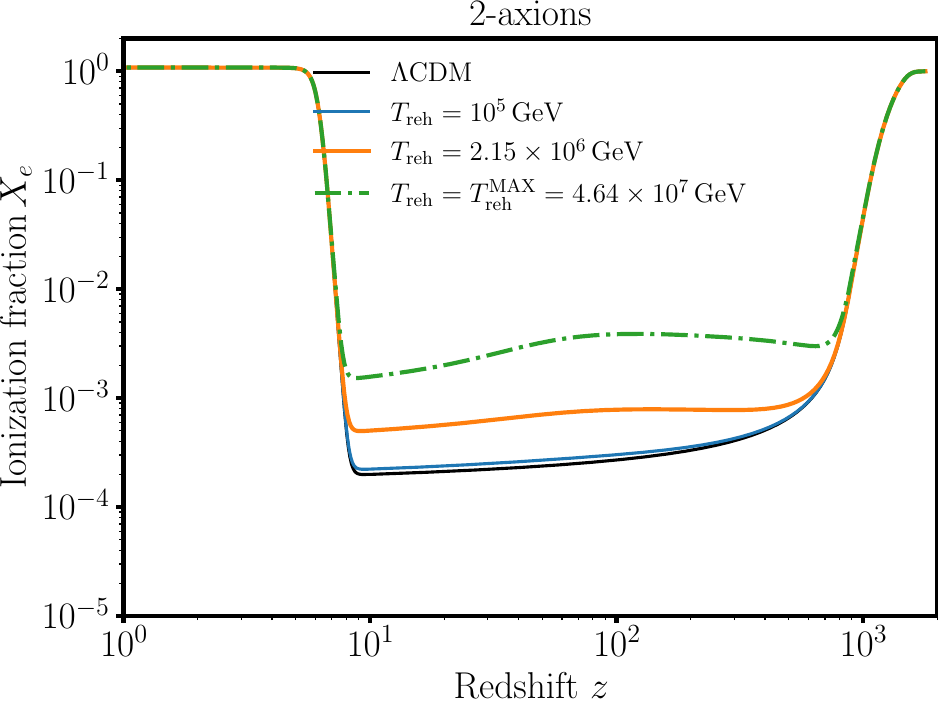 }
    \caption{\textbf{Multiple axions case.} Ionization history for a specific string axion model in an ensemble with $h^{1,1}=50$, containing two axions within the relevant mass window. The axion parameters are: $m_{a}^{(1)}=8.27$\,GeV, $g_{a\gamma\gamma}^{(1)}=1.87\times 10^{-20}\,\mathrm{GeV}^{-1}$, and  $m_{a}^{(2)}=0.72\,\mathrm{GeV}$, $g_{a\gamma\gamma}^{(2)}=2.1\times 10^{-20}\,\mathrm{GeV}^{-1}$. Different line styles correspond to the reheat temperatures in the legend.  }
    \label{fig:axiondecay}
\end{figure}

Figure~\ref{fig:axiondecay} illustrates the ionization history $X_e(z)$ resulting from a representative string axion model within the $h^{1,1} = 50$ ensemble. Details for the computations are similar to those outlined for Fig.~\ref{fig:Xez}. The selected string axion model includes two axions within the mass range relevant for contributions to the CMB optical depth: one with mass $m_{a}^{(1)} = 8.27\,\mathrm{GeV}$ and photon coupling $g_{a\gamma\gamma}^{(1)} = 1.87 \times 10^{-20}\,\mathrm{GeV}^{-1}$, and a second with $m_{a}^{(2)} = 0.72\,\mathrm{GeV}$ and $g_{a\gamma\gamma}^{(2)} = 2.1 \times 10^{-20}\,\mathrm{GeV}^{-1}$. These axions decay during the early Universe, injecting energy that modifies the evolution of the free electron fraction before $z \approx 10$. 
The resulting ionization history departs from the standard $\Lambda$CDM scenario, potentially imprinting on the CMB polarization through the enhanced $\tau_{\mathrm{highz}}$. Results are shown for different values of the reheating temperature, up to the maximally allowed one, $T_{\rm reh}^{\rm MAX}$, defined as the value at which the computed optical depth exceeds the $2\sigma$ exclusion threshold from our $\tau_{\mathrm{highz}}$ analysis in Paper~I~\cite{Cheng:2025cmb}; see Fig.~\ref{fig:tau_zc} below.

\section{The high-\texorpdfstring{$z$}{z} optical depth}
\label{sec:highztau}

In our companion Paper~I~\cite{Cheng:2025cmb}, we adopt a model-independent approach to constrain the optical depth $\tau_{\mathrm{reio}}$:
\begin{equation}
    \tau_{\mathrm{reio}} = \int^{z_{\mathrm{max}}}_0 \sigma_T\,n_e(z)\,\frac{\mathrm{d}z}{(1+z)H(z)},
\label{eq:tau_reio}
\end{equation}
where we fix $z_{\mathrm{max}} = 800$. To better understand the contributions to the total optical depth from different epochs of reionization, we decompose $\tau_{\mathrm{reio}}$ into two components: a low-redshift contribution $\tau_{\mathrm{lowz}}$ and a high-redshift contribution $\tau_{\mathrm{highz}}$. This decomposition is particularly useful for distinguishing between standard astrophysical reionization and potential exotic energy injection scenarios that could ionize the Universe at much earlier times. Following Eq.~\eqref{eq:tau_reio}, we define the optical depth at low and high redshifts in terms of a critical redshift $z_c$:
\begin{equation}
\label{eq:tau_lowhighz}
\begin{split}
    \tau_{\mathrm{lowz}} &= \int^{z_c}_0 \,\sigma_T\,n_e(z)\,\frac{\mathrm{d}z}{(1+z)H(z)},\\
    \tau_{\mathrm{highz}} &= \int^{z_{\mathrm{max}}}_{z_c}\,\sigma_T\,n_e(z) \frac{\mathrm{d}z}{(1+z)H(z)}\,,
\end{split}
\end{equation}
so that $\tau_{\mathrm{reio}} = \tau_{\mathrm{lowz}} + \tau_{\mathrm{highz}}$.

\begin{figure}[htbp]
    \centering
    \includegraphics[width=\linewidth]{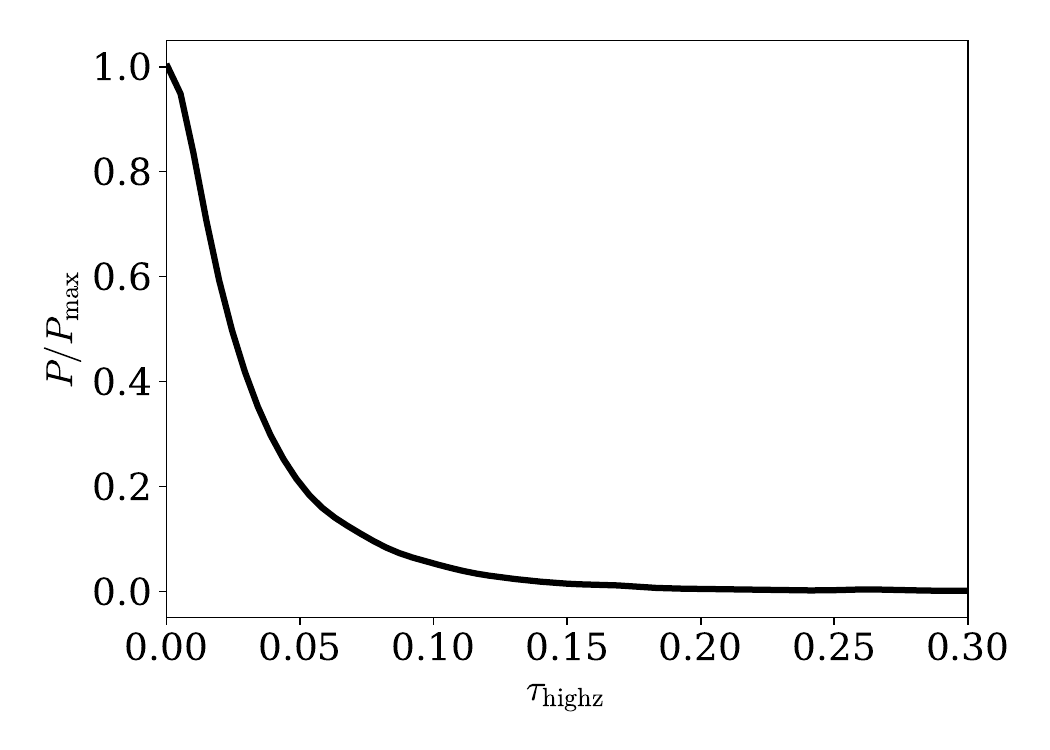}
    \caption{Normalized posterior distribution, $P/P_{\mathrm{max}}$, for the high–redshift contribution to the optical depth, for the critical redshift $z_c=30$ defining the separation in Eq.~\eqref{eq:tau_lowhighz}. Results are based on the \textit{Planck} low-$\ell$ EE dataset alone and the analysis in Paper~I. This posterior distribution gives the upper limit $\tau_{\mathrm{highz}} < 0.111$ at 95\% CL, which we use to set limits on axion models.}
    \label{fig:tau_zc}
\end{figure}

The novelty presented in Paper~I~\cite{Cheng:2025cmb} consists in employing a Gaussian process (GP) method to reconstruct arbitrary reionization histories $X_e(z)$~\cite{Cheng:2025cmb}. Using a Markov Chain Monte Carlo (MCMC) analysis with $\mathcal{O}(20)$ parameters describing the GP, together with the \textit{Planck} low-$\ell$ E-mode EE-only SimAll likelihood,\footnote{The most direct and robust probe of the reionization epoch is provided by the low-$\ell$ E-mode polarization data.} we derive limits on $X_e(z)$ that extend to high redshifts.
These constraints are then compressed into a posterior distribution for the high-redshift optical depth $\tau_{\mathrm{highz}}$. Our reconstruction is nonparametric in $X_e(z)$, with no assumed reionization template such as a \texttt{tanh} transition, but is conditioned on a fixed \textit{Planck} $\Lambda$CDM background. The relevant high-$\ell$ information is incorporated through this conditioning, particularly through the fixed value of $A_s$, which is constrained together with the total optical depth through the combination $A_s e^{-2\tau}$~\cite{Planck:2018vyg}. Within this fixed-background framework, explicitly including the high-$\ell$ likelihood would not provide additional information that cleanly isolates $\tau_{\mathrm{highz}}$. The novelty of our approach is the freedom allowed in the shape of $X_e(z)$, rather than independence from the assumed cosmological background.

Our results yield a constraint $\tau_{\mathrm{lowz}} = 0.0483^{+0.0083}_{-0.0080}$ at 68\% confidence level (CL), and an upper limit of $\tau_{\mathrm{highz}} < 0.111$ at 95\% CL, using $z_c = 30$. Figure~\ref{fig:tau_zc} shows the normalized posterior distribution for the high-redshift contribution to the optical depth, $\tau_{\mathrm{highz}}$, as derived from the \textit{Planck} low-$\ell$ EE polarization data, which serves as our constraint on axion models. The resulting posterior can be used to constrain exotic reionization histories beyond the decaying-axion scenario considered here. The distributions are largely insensitive to the choice of the critical redshift $z_c$, as we have demonstrated in Paper~I~\cite{Cheng:2025cmb}. 
The code used to cast the reionization results can be found at:~\href{https://github.com/Cheng-Hanyu/CLASS_reio_gpr}{github.com/Cheng-Hanyu/CLASS\_reio\_gpr}.

Our GP method extends previous model-independent analyses of the reionization history based on CMB large-angle polarization, including the principal component approach of Ref.~\cite{Heinrich:2016ojb} and the reconstruction methods adopted by the \textit{Planck} collaboration to study the standard low-$z$ reionization history~\cite{Planck:2018nkj}. Our result for $\tau_{\mathrm{lowz}}$ is consistent with their findings. In addition, the extracted posterior for $\tau_{\mathrm{highz}}$ yields robust, model-independent constraints on exotic energy injection. By comparing theoretical predictions for $X_e(z)$ that reproduce the empirical posterior of $\tau_{\mathrm{highz}}$, a wide range of scenarios can be directly tested, as we demonstrate below.\footnote{In Paper~I~\cite{Cheng:2025cmb}, we validate this approach with a direct $\chi^2$ comparison using the \textit{Planck} low-$\ell$ EE-only SimAll likelihood.}

\section{Single axion results}
\label{sec:singleaxion}

Before embarking on our analysis with multiple axions in string theory models, we first present constraints on the parameter space $(m_a, g_{a\gamma\gamma}, T_{\rm reh})$ for a single-axion model. This allows us to benchmark our method against existing results in the literature and to present a new result on the maximum allowed value of $T_{\rm reh}$ across the axion parameter space.

\begin{figure}[htb]
    \centering
    \includegraphics[width=1\linewidth]{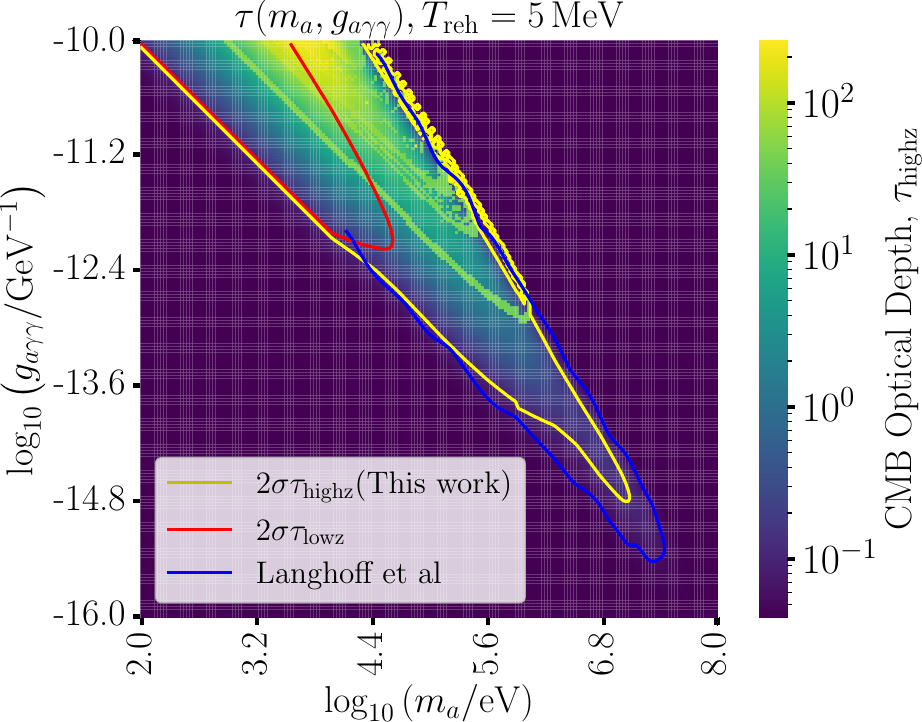}
    \caption{Optical depth $\tau_{\mathrm{highz}}$ (vertical color scaling) as a function of the axion mass (horizontal axis) and axion–photon coupling (vertical axis). Solid lines denote 95\% CL ($2\sigma$) exclusion contours from different analyses. The yellow line shows constraints from our high-redshift optical depth estimation using GP regression results marginalized over $z \in [30, 800]$. The red line represents low-redshift constraints marginalized over $z \in [0, 30]$ from our GP regression analysis, which are consistent with the optical depth estimation from the \textit{Planck} 2018 result~\cite{Planck:2018nkj}. The blue line shows CMB anisotropy constraints from Ref.~\cite{Langhoff:2022bij}, labeled ``Langhoff et al.'' See details about the GP regression and MCMC analysis in Paper~I~\cite{Cheng:2025cmb}.}
    \label{fig:dtaumapTRH5MeV}
\end{figure}

Figure~\ref{fig:dtaumapTRH5MeV} shows the optical depth $\tau_{\mathrm{reio}}$ in the $(m_a, g_{a\gamma\gamma})$ plane for a reheating temperature $T_{\rm reh} = 5\,\mathrm{MeV}$. We show three separate contours for comparison. The yellow contour shows the $2\sigma$ constraint derived from our model-independent analysis based on limits on high-redshift energy injection, labeled $\tau_{\mathrm{highz}}$~\cite{Cheng:2025cmb}. The red contour corresponds to the exclusion derived by comparing the total optical depth $\tau_{\rm reio}$ to the \textit{Planck} constraint, assuming standard low-redshift reionization only, denoted $\tau_{\mathrm{lowz}}$. For reference, the blue contour reproduces the bound from Ref.~\cite{Langhoff:2022bij}. The overall shape of the contours is governed by the interplay between the axion decay lifetime and its freeze-in abundance. The upper-right boundary follows $g_{a\gamma\gamma} \propto m_a^{-3/2}$, corresponding to axions that decay before recombination and do not contribute to the ionization history. The lower-left boundary comes from the suppression of the freeze-in abundance at small masses and couplings, while the strongest constraints arise in the region where the axion abundance is large and the decay occurs within the redshift interval relevant for reionization. Our analysis demonstrates that our model-independent treatment accurately reproduces the limits from previous work~\cite{Langhoff:2022bij, Poulin:2016anj}, while being only marginally more conservative. In contrast, using the full $\tau_{\rm reio}$ constraint without accounting for the standard low-redshift reionization history leads to overly conservative exclusions. 

We have thus validated two key elements of our pipeline: (i) our freeze-in fits~\cite{Jain:2024dtw}, which are consistent with the relic abundance calculations in Ref.~\cite{Langhoff:2022bij}; and (ii) our model-independent CMB analysis of the ionization history, which leads to the $\tau_{\mathrm{highz}}$ posterior and agrees with the results obtained using the ``on-the-spot'' approximation developed in Ref.~\cite{Poulin:2016anj}.

The on-the-spot approximation, relevant for general decaying dark matter scenarios, assumes that: (i) energy from decay is deposited instantaneously into photons; and (ii) the efficiency of the energy deposition follows a simple exponential decay law, given by
\begin{equation}
    f(z) = f_{\rm eff}\, e^{-t(z)\Gamma}\,,
\end{equation}
where $f_{\rm eff}$ depends only on $m_\chi$ and its decay channels, and is defined in~\cite{Poulin:2016anj}.
These models are specified by three parameters: the decay rate $\Gamma$, the DM fraction in the decaying species $\mathcal{F}$, and the mass of the decaying DM component $m_\chi$.

Under the assumptions of the on-the-spot approximation, constraints from CMB anisotropies are obtained by performing MCMC analyses for each set of values $(m_\chi, \Gamma^{-1})$, using flat priors on the six $\Lambda\mathrm{CDM}$ parameters plus $\mathcal{F}$, and deriving upper limits on $\mathcal{F}$ from the data. When applied to axions, this approach yields the fits presented in Ref.~\cite{Langhoff:2022bij}, whose workflow is conceptually and operationally distinct from the one we adopt in this work. Ref.~\cite{Langhoff:2022bij} is based on Ref.~\cite{Poulin:2016anj} and uses more CMB likelihoods than we employ (temperature in addition to polarization), which explains the slightly stronger limit.

\begin{figure}[ht!]
    \centering
    \includegraphics[width=1\linewidth]{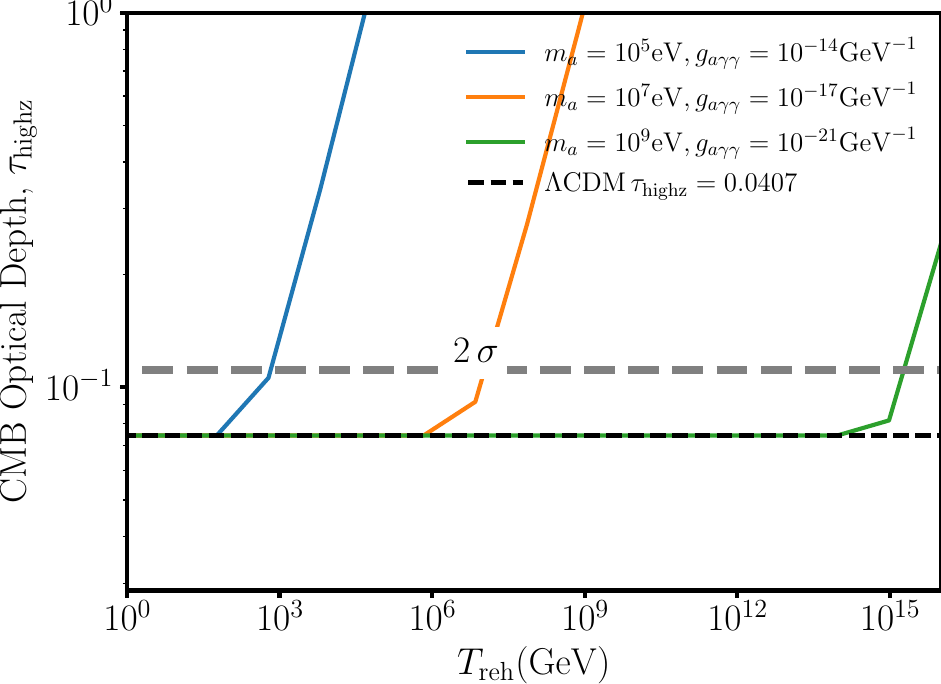}
    \caption{The optical depth $\tau_{\mathrm{highz}}$ as a function of reheating temperature, for a selection of single-axion models with varying mass and photon coupling. The gray dashed line marks the $2\sigma$ upper limit on $\tau_{\mathrm{highz}}$ derived in Paper~I (see also Fig.~\ref{fig:tau_zc}), while the black dashed line shows the $\Lambda$CDM contribution (see main text for details). The intersection of each curve with the gray dashed line defines the maximum reheating temperature $T_{\rm reh}^{\rm MAX}$ allowed for that model.}
    \label{fig:dtauTreh}
\end{figure}

We now apply our method to compute a new result, so far as we are aware, for a single axion: the maximum reheating temperature $T_{\rm reh}^{\rm MAX}$ allowed across the $(m_a, g_{a\gamma\gamma})$ parameter space. For a given axion mass and photon coupling, we calculate the high-redshift optical depth $\tau_{\mathrm{highz}}$ as a function of reheating temperature, using the freeze-in abundance as input. 
Figure~\ref{fig:dtauTreh} shows this dependence for a representative selection of single-axion models, as indicated in the legend. The gray dashed line indicates the $2\sigma$ upper bound on $\tau_{\mathrm{highz}}$ derived in Paper~I (and shown in Fig.~\ref{fig:tau_zc}), which sets the exclusion threshold. As $T_{\rm reh}$ increases, the freeze-in abundance grows, resulting in an increase in $\tau_{\mathrm{highz}}$. For each model, we identify the point at which $\tau_{\mathrm{highz}}$ crosses the exclusion threshold: this defines the corresponding maximum reheating temperature $T_{\rm reh}^{\rm MAX}$ allowed by current data. 
For comparison, the black dashed line shows the value $\tau_{\mathrm{highz}} = 0.0407$, obtained from the baseline $\Lambda$CDM scenario with $z_c = 30$ in Eq.~\eqref{eq:tau_lowhighz}.

\begin{figure}[ht!]
    \centering
    \includegraphics[width=1\linewidth]{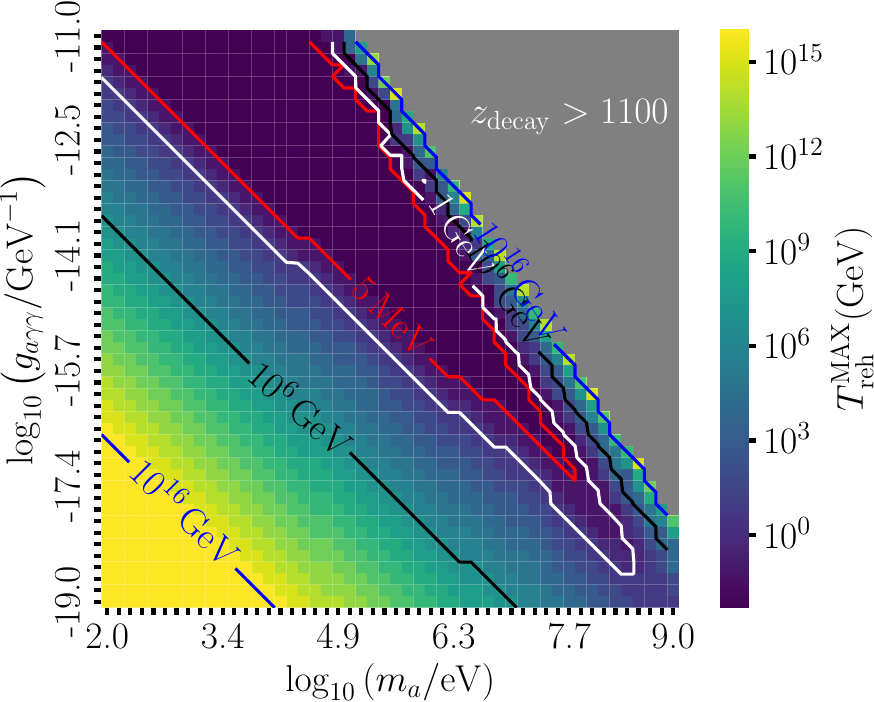}
    \caption{The maximum reheating temperature $T_{\rm reh}^{\rm MAX}$ allowed by CMB optical depth constraints, shown as a function of $m_a$ and $g_{a\gamma\gamma}$ for a single axion-like particle. Solid contours indicate constant values of $T_{\rm reh}^{\rm MAX}$: 5\,MeV (red), 1\,GeV (white), $10^6$\,GeV (black), and $10^{16}$\,GeV (blue). The lower-left region of the plot is compatible with high reheating temperatures, including values near $\sim 10^{16}\,\mathrm{GeV}$, consistent with expectations from inflationary gravitational wave constraints~\cite{Domcke:2015iaa}. The gray upper-right region is unconstrained by our analysis, as the axion decays before recombination and thus does not contribute to the high-redshift optical depth (see Ref.~\cite{Balazs:2022tjl} for constraints in this region).}
    \label{fig:max_T_single_ALP}
\end{figure}

Figure~\ref{fig:max_T_single_ALP} shows the maximum reheating temperature $T_{\rm reh}^{\rm MAX}$ allowed by CMB optical depth constraints, mapped over the $(m_a, g_{a\gamma\gamma})$ parameter space for a single axion-like particle. For each point in this plane, we compute the freeze-in abundance and the resulting high-redshift optical depth $\tau_{\mathrm{highz}}$, identifying the highest reheating temperature for which the predicted $\tau_{\mathrm{highz}}$ remains below the $2\sigma$ exclusion threshold.
Our procedure smoothly interpolates between two limiting cases: the low-reheating bound from Ref.~\cite{Langhoff:2022bij}, evaluated at $T_{\rm reh} = T_{\rm BBN}$, and the fully thermalized limit from Refs.~\cite{Cadamuro:2010cz, Cadamuro:2011fd}, corresponding to $T_{\rm reh} \sim 10^{16}\,\mathrm{GeV}$. Contours of constant $T_{\rm reh}^{\rm MAX}$ are overlaid to indicate the energy scale at which each region of parameter space becomes excluded. 
This figure thus serves as a direct map between axion properties and the maximum cosmologically allowed reheating temperature. Notably, the lower-right region of the parameter space permits high-scale reheating, in agreement with constraints from inflationary gravitational wave backgrounds~\cite{Domcke:2015iaa}, while the upper-right region is unconstrained in our analysis, as axion decays occur before recombination and leave no imprint on the ionization history and $T_{\rm reh}^{\rm MAX}$ is always smaller than the freeze-out temperature $T_{\rm{fo}}$ in the parameter space we are interested in, which guarantees that our freeze-in scenario is valid throughout this work.

\section{Constraints on reheating in string theory}
\label{sec:stringconstraints}

\begin{figure}[ht!]
    \centering
    \includegraphics[width=1.0\linewidth]{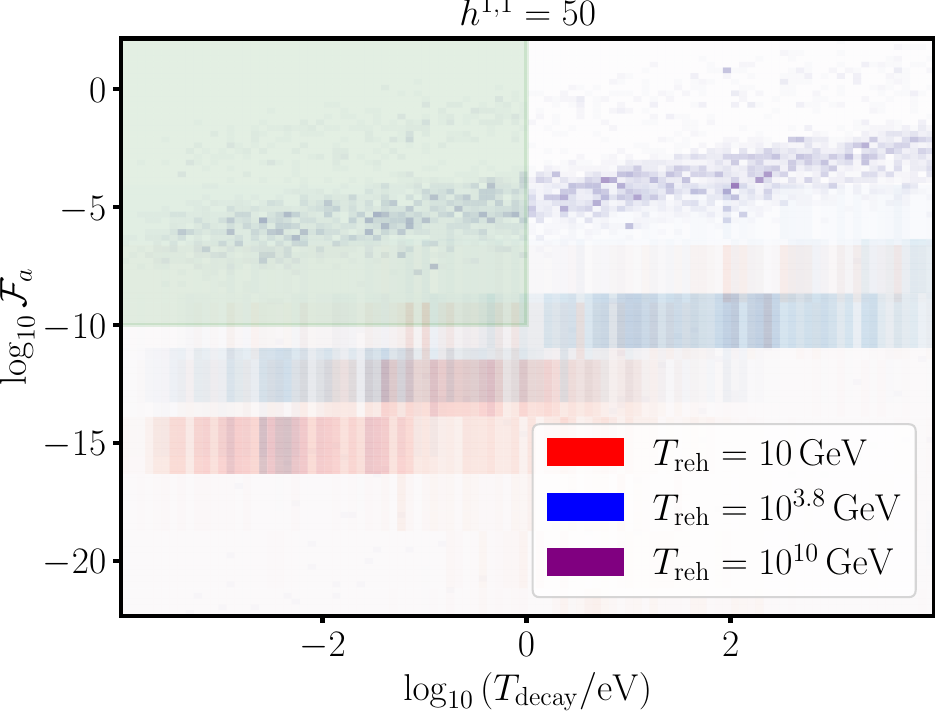}
    \caption{2D histogram of the decay temperature $T_{\rm decay}$ and freeze-in relic abundance fraction $\mathcal{F}_a$ for string axions within the $h^{1,1} = 50$ ensemble. Different color bins correspond to different reheating temperatures. The green shaded region shows the parameter space excluded by existing CMB optical depth constraints on single axion decays~\cite{Poulin:2016anj, Langhoff:2022bij}. The correlation between $\mathcal{F}_a$ and $T_{\rm decay}$ arises from the dependence of the freeze-in abundance and decay rates on the axion parameters $m_a$ and $g_{a\gamma\gamma}$, as given in Eqs.~\eqref{eq:Gamma} and~\eqref{eq:primakoff}. As the reheating temperature increases, the freeze-in abundance grows, shifting models toward this excluded region.}
    \label{fig:2D_Tdec_abundance}
\end{figure}

We now turn to a quantitative assessment of the constraints that can be placed within the string axiverse. We generate ensembles of models as described above, using the methodology of Ref.~\cite{Gendler:2023kjt}. Figure~\ref{fig:2D_Tdec_abundance} shows the distribution of axions in the plane of decay temperature $T_{\rm decay}$ and freeze-in relic abundance fraction prior to decay, $\mathcal{F}_a$, for the ensemble with $h^{1,1} = 50$. This representation provides a mapping between microscopic axion parameters and their impact on the ionization history. Existing limits on single axion decays from the CMB optical depth exclude regions with sufficiently late decay, $T_{\rm decay} \lesssim 1\,\mathrm{eV}$, and large abundance $\mathcal{F}_a \gtrsim 10^{-10}$~\cite{Poulin:2016anj, Langhoff:2022bij}, shown as the green shaded region in Fig.~\ref{fig:2D_Tdec_abundance}. From the figure, it appears that a subset of models within the ensemble enter this excluded region, particularly for larger reheating temperatures $T_{\rm reh} \gtrsim 10^{10}\,\mathrm{GeV}$~\cite{Gendler:2023kjt}. This reflects the fact that increasing $T_{\rm reh}$ enhances the freeze-in abundance, shifting models vertically toward larger $\mathcal{F}_a$ while leaving the decay temperature largely unchanged. The figure shows a potential tension between high reheating temperatures and CMB constraints: models with sufficiently large freeze-in abundance and late decay can produce excessive energy injection during the epoch relevant for reionization. In the following section, we quantify this tension using our full multi-axion analysis.

Each model contains many axions, spanning a broad range of masses and decay constants. We preprocess the ensembles to ensure both the accuracy of the numerical results and computational feasibility. Specifically, we define a mass window from $10^5\,\mathrm{eV}$ to $10^{10}\,\mathrm{eV}$, 
corresponding to axions that decay in the redshift interval relevant for CMB constraints on reionization, and select at most 10 axions per model within this window for further analysis. This selection is justified by the expected axion mass hierarchies and the approximately uniform coupling to photons (for axions above the ``light threshold'' mass, which typically holds for high-mass axions), which together ensure that only a small number of axions significantly contribute to the CMB optical depth. This preprocessing step reduces the number of models that must be evaluated. Empirically, the number of axions within this window rarely exceeds 10 even prior to truncation, so this selection does not bias the physical results.

\begin{figure}[htbp]
    \centering
    \includegraphics[width=1.0\linewidth]{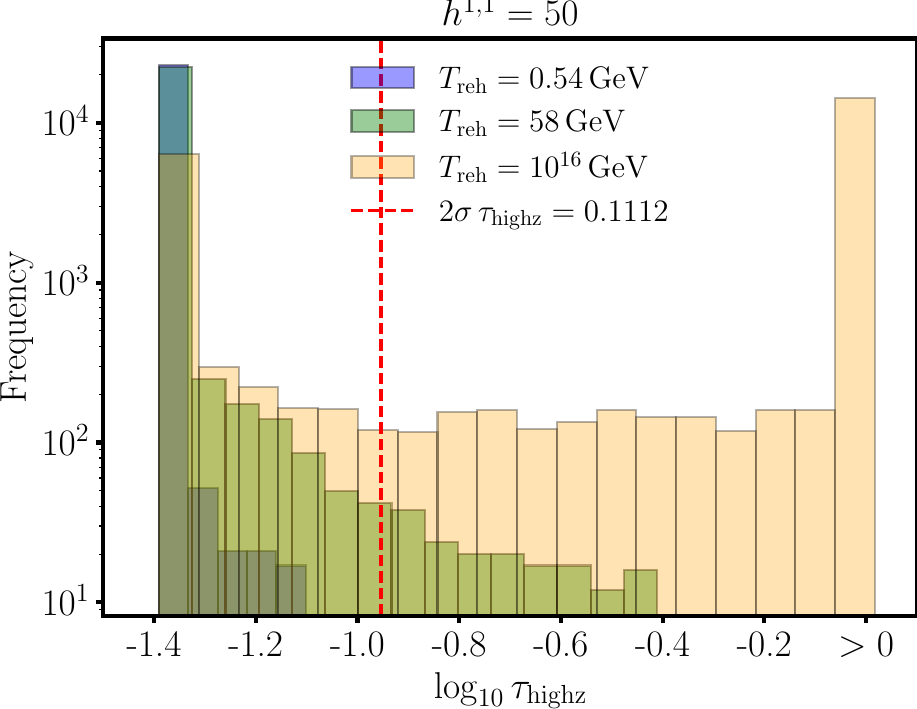}
    \caption{Distribution of the high-redshift optical depth $\tau_{\mathrm{highz}}$ for string axion models with $h^{1,1} = 50$. The red dashed line indicates the $2\sigma$ exclusion threshold, corresponding to $\tau_{\mathrm{highz}} < 0.111$, based on our data analysis as described in Paper~I.}
    \label{fig:taulow}
\end{figure}

For all selected axions in a given model, we then compute the freeze-in abundance and evaluate the optical depth $\tau_{\mathrm{highz}}$ from their decay, using the methods of Sec.~\ref{sec:reionization}. To avoid numerical instabilities from unphysically large values of the optical depth during the calculations, we impose an upper limit $\tau_{\mathrm{highz}} \leq 1$ which lies well above the observational exclusion threshold. Figure~\ref{fig:taulow} shows the distribution of $\tau_{\mathrm{highz}}$ for models with $h^{1,1} = 50$, for different reheating temperatures. The red dashed line indicates the $2\sigma$ exclusion threshold derived from our posterior analysis of $\tau_{\mathrm{highz}}$. For each model, we identify the maximum reheating temperature $T_{\rm reh}^{\rm MAX}$ at which the model remains consistent with this threshold.

\begin{figure}[htbp]
    \centering
    \includegraphics[width=\linewidth]{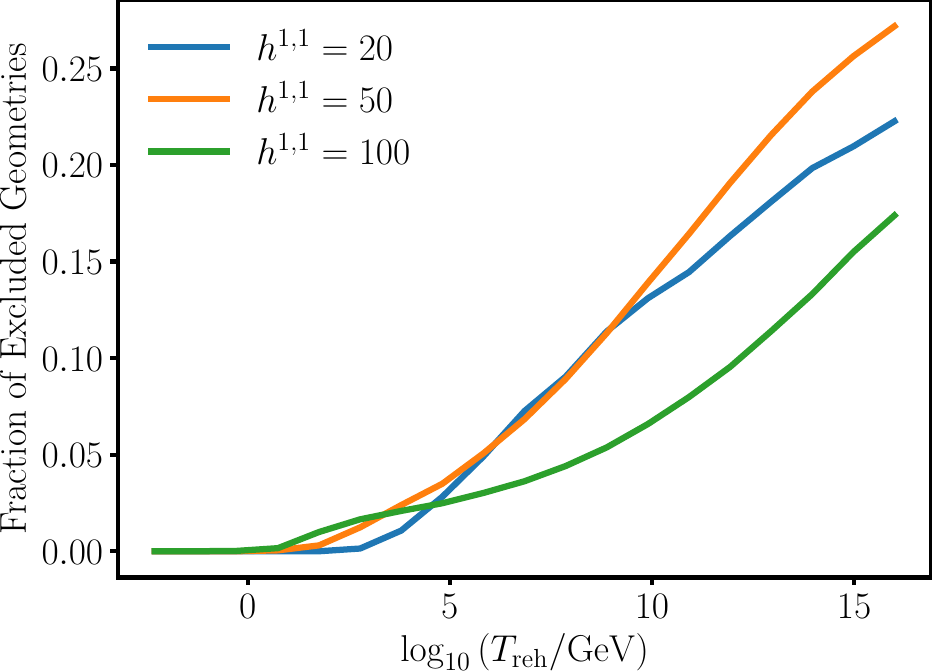}
    \caption{Fraction of excluded string theory models at 95\% CL as a function of the reheating temperature $T_{\mathrm{reh}}$ for different Hodge numbers $h^{1,1}$. Models are constructed as described in Sec.~\ref{sec:axioncosmology}, following Ref.~\cite{Gendler:2023kjt}, using \textsc{CYTools}~\cite{Demirtas:2022hqf}. The relic abundance is computed following Ref.~\cite{Jain:2024dtw}. The full ionization history is evaluated using the methods described in this paper, in conjunction with \textsc{DarkHistory}~\cite{Liu:2019bbm}. Constraints are derived from the CMB optical depth, following the model-independent analysis presented in Paper~I~\cite{Cheng:2025cmb}.}
    \label{fig:exclusion_reheat}
\end{figure}

Our final results are displayed in Fig.~\ref{fig:exclusion_reheat}, where we show the fraction of excluded string theory models in our ensemble as a function of the reheating temperature $T_{\rm reh}$ for Hodge numbers $h^{1,1}=20,50,100$. For $h^{1,1}=20,50$, we find that up to 25\% of models are excluded at the largest value of $T_{\rm reh} \approx 10^{16}\,\text{GeV}$, and around 15\% of models are excluded for $T_{\rm reh} \gtrsim 10^{10}\,\text{GeV}$. At $h^{1,1}=100$, the constraints are weaker. While the maximum axion-photon coupling increases with $h^{1,1}$, the corresponding axion mass spectrum is shifted toward smaller values due to the effects of ``kinetic isolation'' and the ``light threshold'' identified in Ref.~\cite{Gendler:2023kjt}. These effects arise from the structure of the K\"ahler metric, which increasingly suppress axion masses as the overall volume grows. As a result, many of the more strongly coupled axions decay either too early or too late to contribute to the optical depth in the relevant redshift range. This behavior reflects a competition between the increasing multiplicity and coupling strength of axions, and the decreasing fraction of axions with masses in the phenomenologically relevant range. The results in Fig.~\ref{fig:exclusion_reheat} use the more accurate K\"ahler cone $\mathcal{K}_\cup$, which improves over the computation in Ref.~\cite{Gendler:2023kjt}, where $\mathcal{K}_V$ is instead used. Constraints using $\mathcal{K}_\cup$, where it can be computed with present methods, are stronger compared to $\mathcal{K}_V$ due to the presence of more small divisors in $\mathcal{K}_\cup$. We expect, furthermore, that the accuracy of the K\"ahler cone becomes more important at large $h^{1,1}$, and thus further work is required to place explicit constraints in this important part of the landscape~\cite{Demirtas:2020dbm}.

\section{Conclusions}
\label{sec:conclusions}

We have developed a robust and efficient pipeline to evaluate the impact of decaying string axions on the high-redshift CMB optical depth, $\tau_{\mathrm{highz}}$, which is a sensitive probe of light decaying relics. The string axiverse effective field theories derived from type IIB Calabi-Yau orientifold compactifications contain $h^{1,1}$ axions. We constructed explicit models using \textsc{CYTools}~\cite{Demirtas:2022hqf}, following Ref.~\cite{Gendler:2023kjt}, at $h^{1,1}=20,50,100$. For each model and axion, we computed the freeze-in abundance~\cite{Jain:2024dtw} and, using the methods developed in the current paper and a modified version of \textsc{DarkHistory}~\cite{Liu:2019bbm}, calculated the full ionization history $X_e(z)$ for simultaneous decays of all axions. We then derived ${\mathrm{highz}}$ and, following Paper~I~\cite{Cheng:2025cmb}, imposed a conservative exclusion threshold of $\tau_{\mathrm{highz}} < 0.111$ (95\% CL). This allowed us to determine the fraction of excluded models in the axiverse ensemble as a function of the reheating temperature. At $h^{1,1} = 20, 50, 100$, we found that 15\%, 15\%, and 10\% of models prefer a reheating temperature $T_{\rm reh} \lesssim 10^{10}\,\text{GeV}$ at 95\% CL using our model-independent test. For the non-monotonic behavior of the exclusion fraction curves for different Hodge numbers, there are three competing effects that likely underlie that. Increasing $h^{1,1}$ raises the maximal axion-photon coupling and the number of axions with a chance to fall in the mass range relevant for reionization, strengthening the constraint from $h^{1,1}=20$ to $50$. However, in larger $h^{1,1}$, the maximal axion mass, which is set roughly by the KK scale, decreases, shifting the mass distribution to lower values and reducing this probability. A systematic, high-statistics study at large $h^{1,1}$ is therefore needed to clarify this trend in the future.

We have focused on axion-photon interactions and neglected axion couplings to fermions, which are assumed to be subdominant in the region of parameter space considered. Extensions of this analysis could include axion-fermion couplings, which would modify the production mechanisms and open additional decay channels into Standard Model particles, which later branch into electrons. In fact, significant progress has been made in recent years on fermion-driven axion production~\cite{Arias-Aragon:2020shv, Caloni:2022uya, DEramo:2022nvb, DEramo:2025jsb, Barbieri:2025moq, Barbieri:2026ewj}, making such an extension viable once axion-fermion couplings can be computed in ensembles of compactifications of type IIB string theory. Note, however, that these couplings are model-dependent and may be suppressed compared to axion-photon interactions~\cite{Cicoli:2012sz, Halverson:2017deq}. In particular, they depend on a computation of matter couplings in the K\"ahler potential. Nevertheless, axion couplings to fermion fields are generically induced at loop level from the axion-photon coupling.

In the present work, we have used a single observable, namely the high-$z$ CMB optical depth upper limit derived in Paper~I~\cite{Cheng:2025cmb}, to exclude a small but significant fraction of string theory models with high-temperature reheating. The same underlying physics of the axion, freeze-in production followed by decay, also affects several other observables, including X-ray line searches, big bang nucleosynthesis, and CMB spectral distortions~\cite{Langhoff:2022bij,Balazs:2022tjl}, all of which could be used to further constrain the axiverse~\cite{Gendler:2023kjt}. However, just as we here developed a new calculation of the ionization state of the Universe, including multi-axion energy injection, to properly compute the CMB optical depth constraint, so exploiting these additional datasets will require new calculations that include multi-axion decays. Incorporating additional observables alongside the CMB optical depth will greatly extend the range of string theory models that can be constrained in models with high-temperature reheating.

\vspace{0.5cm}
\acknowledgments{
HC, ZY, LV acknowledge support by the National Natural Science Foundation of China (NSFC) through the grant No.\ 12350610240 ``Astrophysical Axion Laboratories''. This publication is based upon work from the COST Actions ``COSMIC WISPers'' (CA21106) and ``Addressing observational tensions in cosmology with systematics and fundamental physics (CosmoVerse)'' (CA21136), both supported by COST (European Cooperation in Science and Technology). HC, ZY acknowledge the support of the China Scholarship Council program (Project ID:202406230341 and 202406230357). EDV is supported by a Royal Society Dorothy Hodgkin Research Fellowship. DJEM is supported by an Ernest Rutherford Fellowship from the Science and Technologies Facilities Council, STFC (Grant No. ST/T004037/1), an STFC consolidator grant (Grant No. ST/X000753/1) and by a Leverhulme Trust Research Project (Grant No. RPG-2022-145). NG is supported in part by a grant from the Simons Foundation (602883,CV) and
gifts from the DellaPietra Foundation. LV also acknowledges support by Istituto Nazionale di Fisica Nucleare (INFN) through the Commissione Scientifica Nazionale 4 (CSN4) Iniziativa Specifica ``Quantum Universe'' (QGSKY). LV thanks the Tsung-Dao Lee Institute for hospitality during the final stages of this work. We are grateful to Mudit Jain for providing assistance with the fits used for the Freeze-in calculation, to Sebastian Vander Ploeg Fallon for investigating the dependence of axion spectra on different approximations to the K\"ahler cone, to Miguel Escudero for correspondence about the baryon temperature, and Liam McAllister and Jakob Moritz for discussions. This work made use of the open source software \textsc{matplotlib}~\cite{2007CSE.....9...90H}, \textsc{numpy}~\cite{2020Natur.585..357H}, and \textsc{scipy}~\cite{2020NatMe..17..261V}.}

\bibliographystyle{apsrev4-1}
\bibliography{references.bib}

\end{document}